\documentclass[sigconf, nonacm]{acmart}
\usepackage{caption}
\usepackage{subcaption}
\usepackage{xcolor}
\newcommand\vldbdoi{XX.XX/XXX.XX}

\newcommand\vldbavailabilityurl{URL_TO_YOUR_ARTIFACTS}
 
\begin{document}
\title{The self-organization of selfishness: Reinforcement Learning shows how selfish behavior can emerge from agent-environment interaction dynamics}

\author{Aamir Sahil Chandroth}
\affiliation{%
  \institution{Learning Sciences Research Group, Homi Bhaba Center for Science Education, Tata institute of Fundamental Research}
  \streetaddress{Anushakti Nagar}
  \city{Mumbai}
  \state{India}
  \postcode{400088}
}
\email{aamir@hbcse.tifr.res.in}

\author{Nithya Ramakrishnan}
\affiliation{%
  \institution{Institute of Bioinformatics and Applied Biotechnology}
  \streetaddress{Electronics City Phase I}
  \city{Bengaluru}
  \state{India}
  \postcode{560100}
}
\email{nithya@ibab.ac.in}


\author{Sanjay Chandrasekharan}
\affiliation{%
  \institution{Learning Sciences Research Group, Homi Bhaba Center for Science Education, Tata Institute of Fundamental Research}
  \streetaddress{Anushakti Nagar}
  \city{Mumbai}
  \state{India}
  \postcode{400088}
}
\email{sanjay@hbcse.tifr.res.in }

\begin{abstract}
In biological communities that use signaling structures for complex coordination, free-riding agents emerge. Such agents do not contribute to the community resources (signals), but exploit them. Most models of such 'selfish' behavior propose that free-riding evolves through mutation and selection. Across many generations, the mutation -- which is considered to create a stable trait -- spreads through the population. This process can lead to a version of the 'Tragedy of the Commons', where the community's coordination resource gets fully depleted or deteriorated. In contrast to this evolutionary view, we present a reinforcement learning model, which shows that both signaling-based coordination and free-riding behavior can emerge within a generation, through learning based on energy minimisation. Further, we show that two types of free-riding can emerge in such systems, and both of these are not stable traits, but dynamic 'coagulations' of agent-environment interactions. Our model thus shows how different kinds of selfish behavior can emerge through self-organization within a lifetime. Further, we  suggest that the view of selfishness as a stable trait presumes a genetic model based on mutations. We conclude with a discussion of some possible social and policy implications of our model.

\end{abstract}

\maketitle



\section{Introduction}
Complex coordination based on the generation and use of signaling structures (stigmergy) is common in many biological systems\cite{bonabeau1997self, couzin2002collective, theraulaz1999brief,torney2011signalling}, particularly microbes (e.g. quorum sensing) and insects (e.g. pheromone trails). Such coordination behaviors, based on the formation and maintenance of community-level signal structures, help optimize the performance of the community as a whole. Given this group-level opimization effect, such behaviors are usually considered to evolve through random mutations, and across generations\cite{deneubourg1989blind, emerson1958evolution}.

However, our previous modeling work\cite{chandrasekharan2007origin,shinde2019recombinant}, based on reinforcement learning (RL), has shown that this kind of complex coordination behavior and community-level signal structures can also emerge within a single generation, based on individual agents lowering their energy load. In recent work by OpenAI, this modeling approach has been extended, to show how the use of physical tools -- and not just symbolic tools like signals -- can also emerge through reinforcement learning\cite{https://doi.org/10.48550/arxiv.1909.07528}. 

In our RL model, communities of minimal and 'reactive' agents -- particle-like entities with very localized and limited actions -- transition from initially random movements to highly coordinated behavior, based on learning to systematically generate and use pheromone trails. The pheromone trails lead to a self-organized community-level signaling system that maximizes foraging performance for all agents. In this model, the transition to the self-organized community behavior emerges through each agent minimizing its energy use. Since energy is used only for actions in this model, the final coordinated activity emerges from all agents moving towards least-action states. 

In biological communities that coordinate based on signaling, 'free riding' behavior can emerge\cite{albanese1985rational}, where some agents do not contribute to the signaling structures\cite{alexander1998riding}, but make use of this community resource. This behavior, when it spreads across the population, can lead to a version of the 'Tragedy of the Commons'\cite{1968,rankin2007tragedy}, where the coordination resource will get fully depleted or deteriorated, leading to the community transitioning back to uncoordinated or random behavior. The free-riding behavior also exists in organisms that form stigmergic structures for coordination, such as microbes \cite{strassmann2000altruism,griffin2004cooperation}. Free-riding behavior also exist in economic systems, where such behavior is termed 'selfish'\cite{hindriks2002free,nielsen2014second}, as it involves exploiting community resources to optimize individual benefit. Such behavior in economic systems can be contained through punishment rules\cite{fehr2000cooperation, eldakar2008selfishness}, and also formal community-level management structures\cite{ostrom1990governing, ostrom2000collective}.

In evolutionary models of signaling, free-riding and the related mechanism of deceptive signaling is considered to evolve through a mutation\cite{emerson1958evolution, bradbury1998principles, carazo2014communication, nakano2013evolution}, which spreads through the population based on the survival advantage it provides. In such models, the system can move to a Tragedy of the Commons situation. The coordination system can return in later generations, but only through another mutation. In contrast to this view, we show how two different kinds of free-riding behavior can emerge without mutation, through reinforcement learning within a generation.

Interestingly, unlike the evolutionary and mutation-based emergence of free-riding behavior, our model shows that such 'selfish' behavior need not be a trait or property of the individual. It can be a distributed property, where the behavior emerges from the agent-environment interactions that lead up to the self-organized coordination. We also show that such a free-riding behavior can avoid the Tragedy of the Commons in a robust fashion, purely through interactions with the changing environment, rather than through punishment rules or formal community structures. 

Our model thus presents a way to understand 'selfish' behavior as a form of emergent, plastic, and self-organized dynamics, rather than the common view of selfishness as a static and constant trait. The following sections present the details of this model.

\section{Model Design}
Our RL model is based on a foraging task\cite{chandrasekharan2007origin}, where many agents wander randomly in an environment. Each agent needs to discover a food source (target) and a nest (home), and make a trip between the target and home locations to get a reward. The multi-agent system that models this task has $N$ number of agents (varied for different experiments), each of which is free to interact with the environment. The agents have the ability to move randomly, and also drop and sense two types of pheromones (home and target pheromones)\cite{bradbury1998principles, deneubourg1989blind}. Agents execute these actions randomly, and each action has a  'tiredness' cost. The RL system (implemented using a Q-Learning algorithm) associated to each agent seeks to minimize this energy cost, by projecting the best action for each environmental state that the agent encounters. This energy-minimization process leads to an emergent and non-random sequence of actions, which generates a thick pheromone trail between the home and target, benefiting all agents. The RL model in each agent seeks to minimize its own energy cost, but this process also optimizes all agents' response to the environment. A detailed description of the model, and the methodology used to train and evaluate it, is below.

The simulation can be accessed through our online interface, \url{https://lsr_lab.gitlab.io/adaptive-behavior-web/external/html/}

\subsection{Q-Learning}

The underlying Q-Learning algorithm we use to optimize each agent's behavior is given by \eqref{eq:Qfunc}. It tries to find the functional relationship between each agent's current state ($S$) and the next action ($A$) it needs to take, to optimize a reward variable ($r$).
\begin{equation}
  Q : S \times A \rightarrow r
  \label{eq:Qfunc}
\end{equation}
The algorithm tracks the state-action relationship using a look-up table (Q-table). The rows of the Q-table corresponds to states, and the columns correspond to the actions that the agents can perform. The entries of this table keep track of the expected long-term reward the agents get on selecting that action for that given state. Suppose that at time $t$ the agent is at state $s_{t}$, then the agent will take the action, $a_{t}$, corresponding to the maximum entry in that row. This is given by \eqref{eq:nextAction}
\begin{equation}
  a_{t} = max(Q_{S=s_{t}})
  \label{eq:nextAction}
\end{equation}
If the reward from this action is $r$, then the Q-table entry is updated as,
\begin{equation}
    Q^{new}(s_{t},a_{t})=Q(s_{t},a_{t})+\alpha(r_{t} + \gamma max Q(s_{t+1},a_{t}) - Q(s_{t},a_{t})
    \label{eq:Qupdate}
\end{equation}
\textmd{where,}
\[\alpha = \textmd{learning rate}\]
\[\gamma = \textmd{discount factor}\]

\subsection{The agents and their world}

The environment consists of a $30$ x $30$ toroidal grid, and the home and target locations are $3$ x $3$ squares within the toroid. The intensity of the pheromone dropped by an agent decreases exponentially over time. Each pheromone that is dropped in a square increases fractionally the pheromone intensity in adjacent squares.
\begin{figure}[h!]
  \centering
  \includegraphics[width=0.45\textwidth]{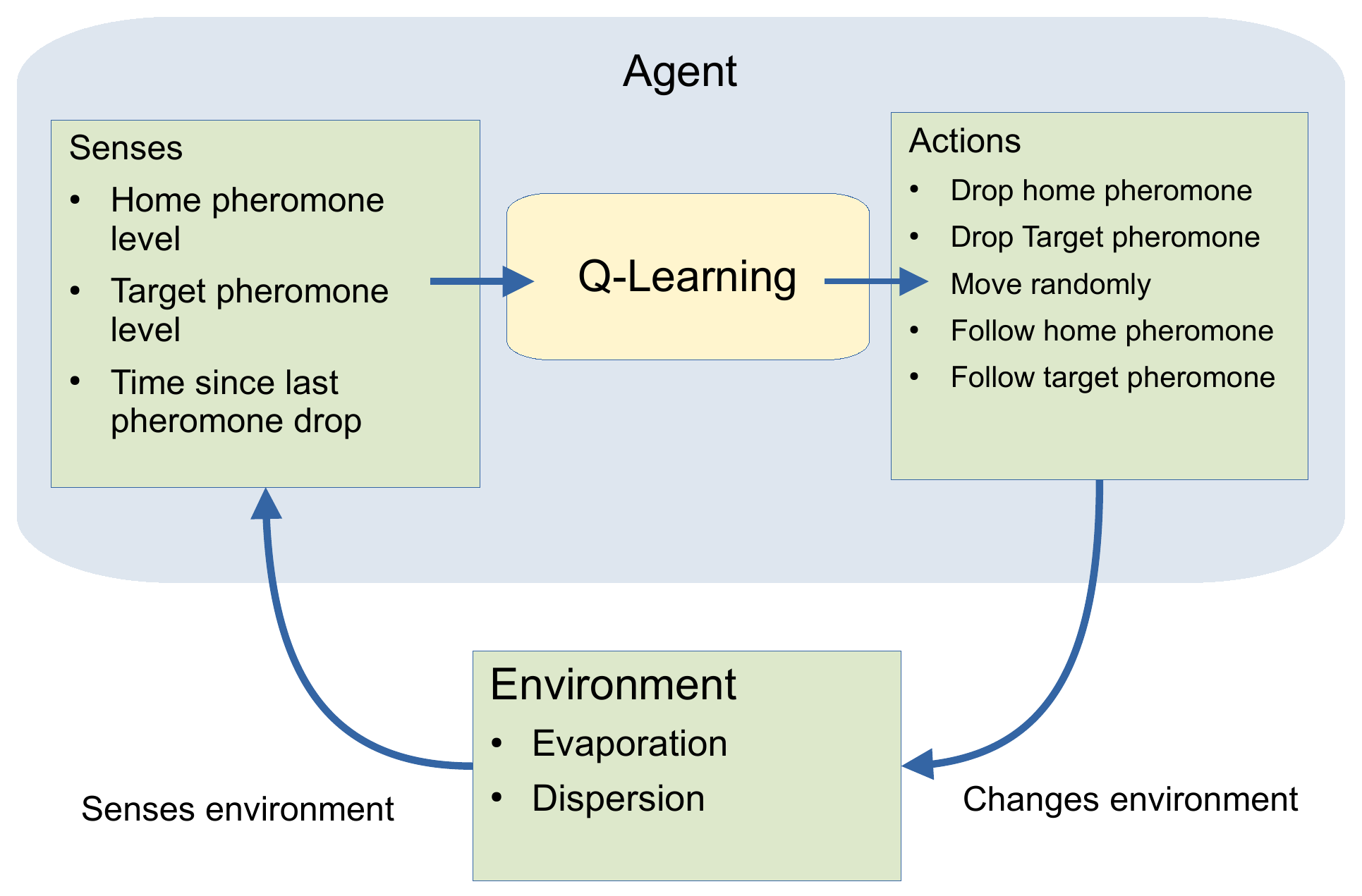}
  \caption{Model design}
  \label{fig:model}
\end{figure}
The agents in the system are equipped with five different actions: moving randomly (straight, diagonal forward-right, diagonal forward-left), moving towards the square with most home pheromone, moving towards the square with most target pheromone(towards the cells that the agent is facing), and generating home or target pheromones. The agents are also equipped with three different external sensors: the level of target likeness and home likeness of the current square, and the time since the last  external structure was generated(the maximum memory being $4$s)\cite{buzsaki2013memory}. This results in a total of $96$ possible states in the Q-table of each agent. These states are tried out randomly by each agent when the simulation starts.

To implement the reinforcement learning system, we designed a reward structure that penalized the agents with -1 point for each action ('tiredness'), and rewarded them with 10 points for completing a trip\cite{rumelhart1986learning} (reaching the home location after visiting the target location). We conducted experiments to evaluate the performance of this RL model under different conditions, including varying the number of agents, the evaporation rate of the pheromones, the exploration rate, and the dispersion rate. Fig \ref{fig:model} and table \ref{tab:parm} summarises the model design.
\begin{table*}[t]
  \caption{Default parameters of the Implementation.\cite{shinde2019recombinant}}
  \label{tab:commands}
  \begin{tabular}{ccl}
    \toprule
    Parameter & Value\\
    \midrule
    \verb|Number of agents| & $40$ \\
    \verb|Learning rate| & $0.2$ \\
    \verb|Discount factor| & $0.5$ \\
    \verb|Exploration rate|  & $0.01$\\
    \verb|Evaporation rate| & $0.99$\\
    \verb|Dispersion rate| & $0.04$\\
    \verb|Total Time| & $100000$\\
    \verb|Time of perturbation | & $50000$\\
    \verb|Number of simulation for averaging| & $50$\\
    \bottomrule
    \label{tab:parm}
  \end{tabular}
\end{table*}
To further evaluate the RL model, we introduced perturbations to the system, and observed their effects. Specifically, we added and removed new agents, and made changes to the environment to see how the agents adapted to these changes. Our aim was to test the robustness and adaptability of the model under different conditions. We provide below a detailed description of the perturbation experiments, including the specific changes made to the environment and the number of agents added or removed. We discuss the results and their implications, particularly for understanding the emergence of collective behavior based on generation and use of symbolic tools (signaling), and the way selfish behavior emerges naturally in such  signaling systems\cite{rankin2007tragedy}.


\section{Results}
\subsection{Emergence of pheromone-generators and free-riders}
Our previous work \cite{chandrasekharan2007origin} showed that the agents could learn to systematically sequence their $96$ possible actions (moving and pheromone drop/sense actions) to generate and follow pheromones, and thus maximize food foraging, individually and also as a community. After a period of time, the initially random actions of all agents (ants) transition to become systematic actions, and this emergent pattern generates an optimal straight pheromone trail between home and target, which allows every agent to maximize food foraging. When this "equilibrium" state is achieved, the number of trips from home to food stays at a constant maximum, for each agent. While some agents continue to generate and drop pheromones during this equilibrium state (labeled as "droppers"), some agents take advantage of the dropped pheromones and enjoy a free-ride with minimum energy cost ("free-riders"). 

Through experiments based on this model, we asked the following questions : Are the behavior of the free-riders and droppers always fixed, or subject to changes in the system? Is there a quorum required for the agents to maintain their state ? Are other environment parameters influencing  the behavior of the agents and the equilibrium journeys? Last but not the least, we ask a hypothetical question: is the past learning sufficient to generate optimal trails, or does the system need to learn constantly?

\subsection{Order of Updation}
In an RL system with multiple agents, for co-operative behavior to emerge, it has been established that the order of actions of the agents is significant\cite{busoniu2008comprehensive, tan1993multi}. Several algorithms can improve the convergence in behavior in multiagent systems \cite{kapetanakis2003reinforcement}. Applied to our ant-foraging model, we observed how the behavior of the agents changed based on the order of actions. If a random action update method is used, where the agents randomly update their individual actions, irrespective of other agents' actions, the behavior of the free-riders and droppers regularly switch between dropping and free-riding. However, in the case of a sequential action update, where the updating of the actions of every individual agent follows a consecutive pattern, more stable behaviors emerge over time.

\subsubsection{Sequential updating}

In the sequential update, at every time step, the actions of agents are updated in the same sequence. This introduces some predictability, and also some structure, to the task environment. However, note that the RL system in each agent needs to detect this global structure to exploit it.

\begin{figure}[h!]
  \centering
  \includegraphics[width=0.45\textwidth]{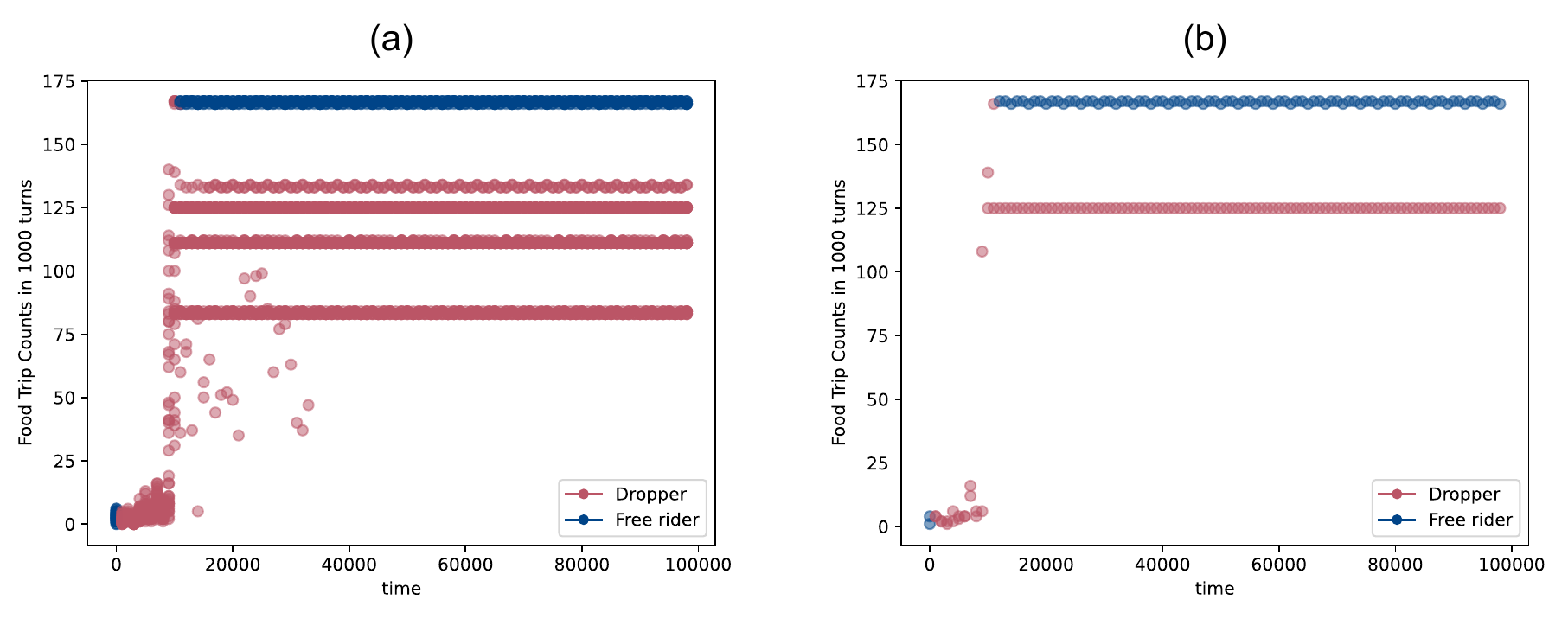}
  \caption{(a) \textmd{\emph{The agents are updated sequentially, leading to the agents behaving as either free-riders or droppers}.} (b) \textmd{\emph{The behaviour of two of the  agents in the system. The dropper remains a dropper and the free-rider remains a free-rider, throughout the simulations.}}}
  \label{fig:seq}
\end{figure}

In sequential updating, we observed that the agents tend to perpetuate either free-riding or dropping behavior. This suggests that all the RL systems (within the agents) do not detect or seek to exploit the global stable structure that is generated by the sequential update. Further, the composition of free-riders and droppers remains  constant over time, as shown in fig \ref{fig:seq}(a). Fig \ref{fig:seq}(b) shows two agents - one a dropper and another a free-rider. As can be seen in the plot, the agents continue to remain in the same behavioral state when the sequential update technique is used.

This behavior suggests that free-riding emerges as a random behavior initially, but it stabilizes because there is enough signaling structures in the environment. In this case, free-riding becomes 'selfish' only when most of the agents move to dropping behavior, and a stable trail emerges because of this steady community contribution. In other words, free riding behavior is 'normal' until a community behavior emerges and this self-organisation leads to overall advantages to the community, including the free-riders. Without an emergent community standard for behavior, and a valuable community resource that is generated through this standardization, there are no 'selfish' behaviors. Any behavior is random -- and thus acceptable -- until the system stabilizes.

\subsubsection{Random updating}
\begin{figure}[h!]
  \centering
  \includegraphics[width=0.45\textwidth]{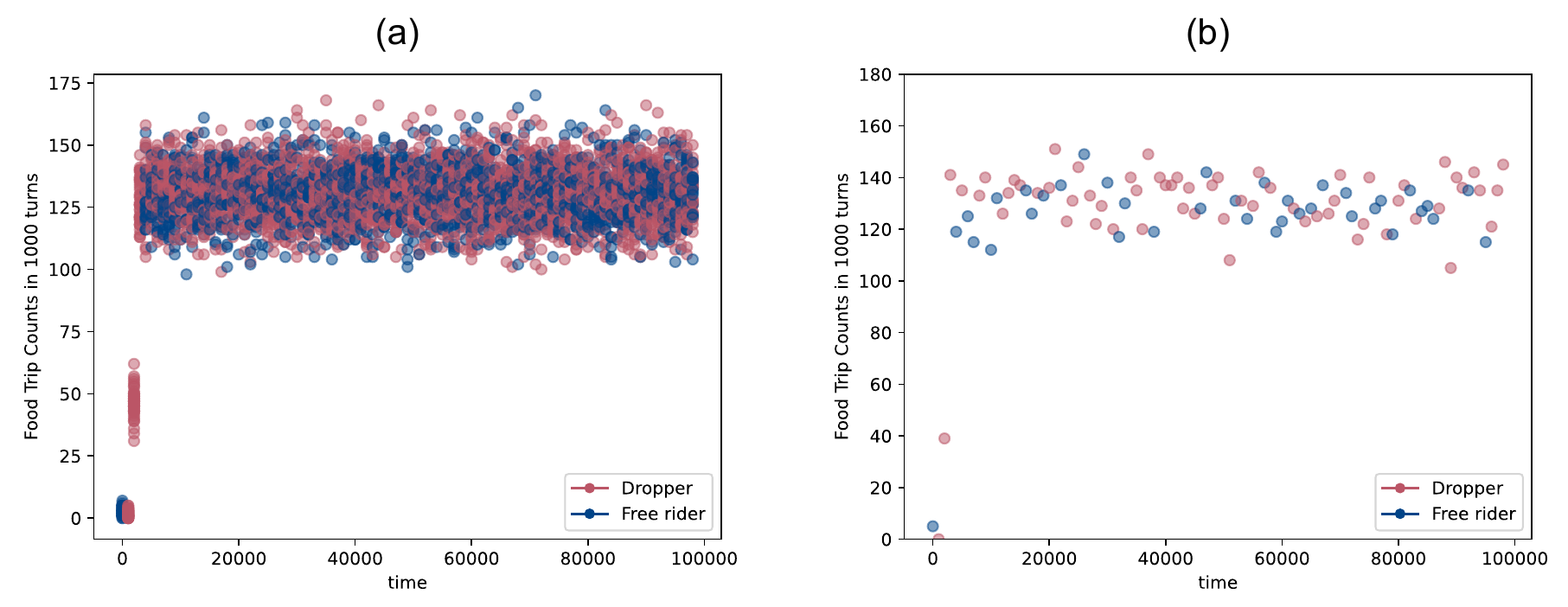}
  \caption{(a) \textmd{\emph{The agents are updated randomly, leading to the agents acting as both free-riders and droppers at different time points}.} (b) \textmd{\emph{The behaviour of one of the agents in the system. It changes its behavior from a dropper to free-rider and vice-versa over time.}}}
  \label{fig:rand}
\end{figure}

In this scenario (fig \ref{fig:rand}), at every time step, the agents update their actions in a random order. In this case, the behavior of the free-riders and droppers regularly switch, indicating that none of the RL systems (in each agent) is able to detect or exploit any stable structure in the environment. This results in a perpetually unstable system, where the composition of free-riders and droppers changes over time.\\
Note that in both sequential and random cases, every agent is acting upon the updated state of the world in a serial manner. To understand how a parallel updating process would affect agent behavior, we ran a further set of experiments. 

\subsubsection{Parallel updating}
\begin{figure}[h!]
  \centering
  \includegraphics[width=0.45\textwidth]{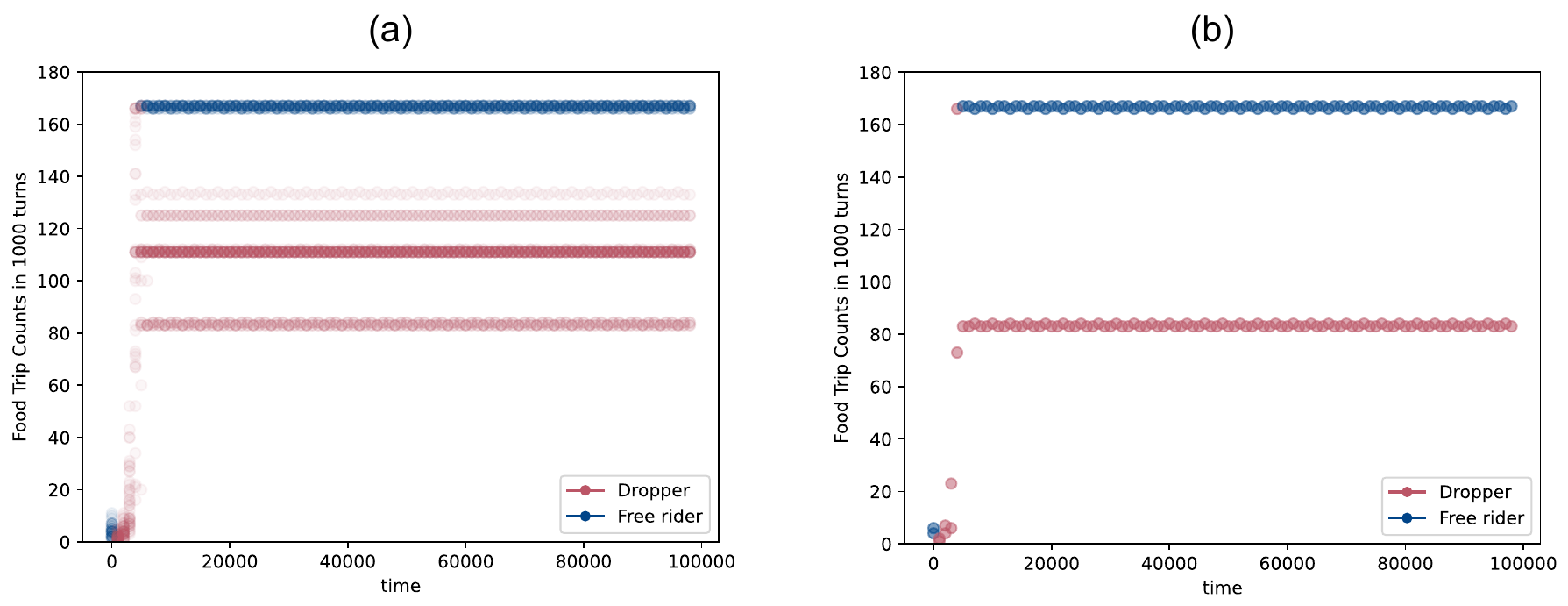}
  \caption{(a) \textmd{\emph{The agents are updated in parallel, leading to the agents acting as either free-riders or droppers similar to the serial case of sequential updation}.} (b) \textmd{\emph{The behaviour of two of the agents in the system. Both the dropper and free-rider stick to their behavior throughout the simulation.}}}
  \label{fig:parallel}
\end{figure}

In this scenario, at every time step, all the agents update their actions by looking at the same state of the world. As can be seen in fig \ref{fig:parallel}, in this case, the free-riders and droppers stick to their respective roles, indicating that their RL systems have found  strategies to utilize the stable structure in the environment. Note that the results are similar to the serial case of sequential updation - with the free-rider and dropper behavior remaining across different scales of time.\\
In the parallel updation case, all the agents are acting upon the same state of the world, collectively. Therefore, the order of updation being random or sequential at a time-step does not have any impact at the end of the iteration.\\
Based on these experiments, we decided to run further simulations based only on the serial sequential updation, as the emergence of free-riding behavior is seen to be independent of the time scales within which the agents are operating (the results for both the sequential(serial) and parallel cases are the same).

\subsection{Death of agents }

The behavior of a complex system can be significantly altered by changes in its components. This is especially true in multi-agent simulations, where the interactions among agents play a crucial role in determining the overall behavior of the system \cite{glaser1997reorganization}. In this experiment, we explored whether the system in equilibrium is perturbed by natural causes. One such example is the death of a few agents. As expected, we saw perturbations in the system when agents were removed (to simulate death) after the system had converged to generating and using pheromones and forming a stable trail.



\begin{figure}[h!]
  \centering
  \includegraphics[width=0.45\textwidth]{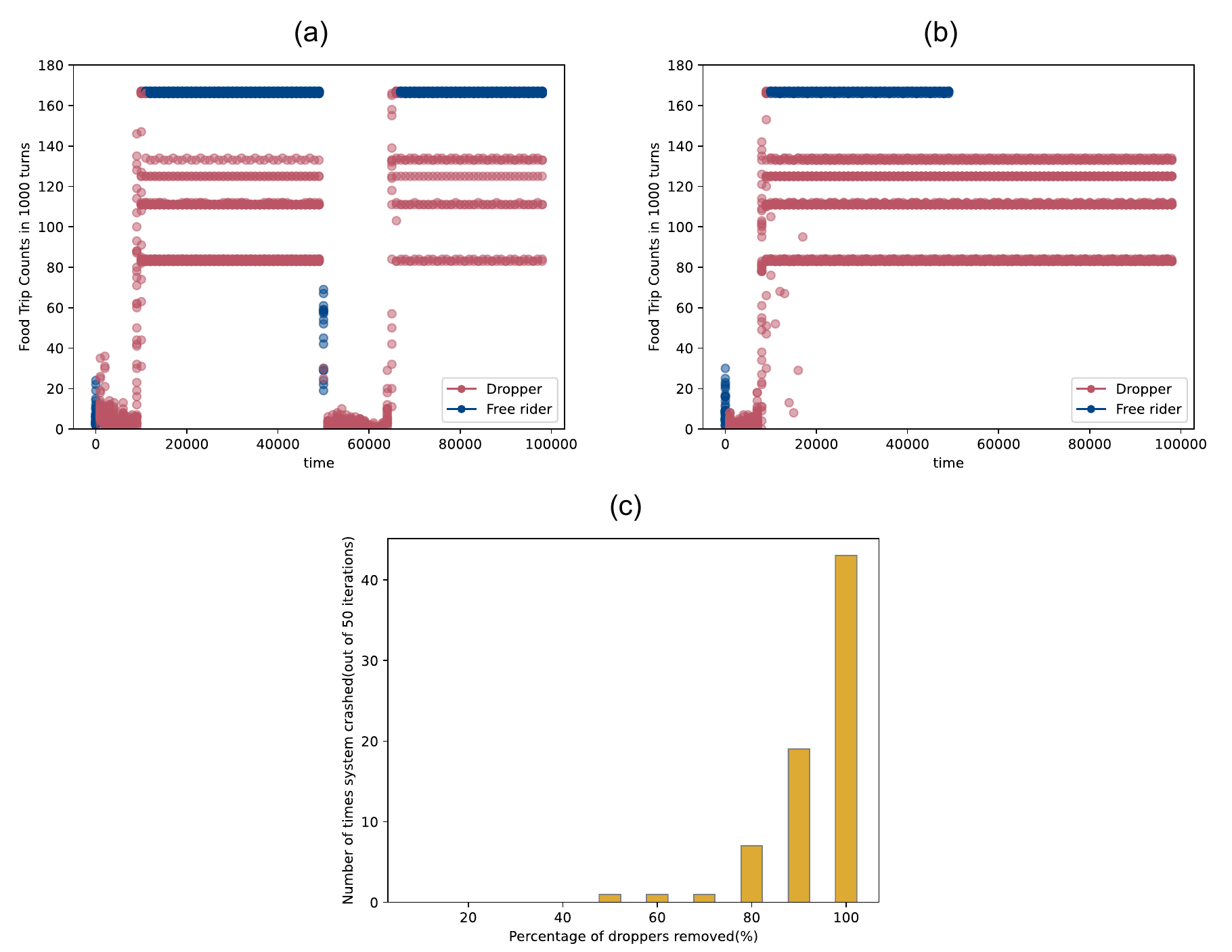}
  \caption{ Death of agents from the system: (a) \textmd{\emph{All the droppers were removed.  In the absence of droppers, some free-riders became droppers.}}(b)\textmd{\emph{ All free-riders were removed.}} (c)\textmd{\emph{ The system starts crashing when close to 50-60\% agents are removed.}}}
  \label{fig:death}
\end{figure}

The first experiment examined the impact of removing one type of agent at $t= 50000$ steps. Two scenarios were considered,  removing (1) droppers and  (2) free-riders. The results are shown in fig \ref{fig:death} (a) and (b). In fig \ref{fig:death}(a), we see that the removal of 20 droppers at $t=50000$ steps causes a perturbation in the system. After $t=62000$, the system comes back to equilibrium. The removal of agents is reflected by a lesser shade of red in the dropper lines as compared to the initial stages. Note that in the absence of any droppers, some of the free-riders change their behavior to become droppers. This indicates free-riding can also emerge as a super-optimization learned by the RL in some agents, based on stable environment structure. Once the stable structure becomes unavailable, these free-riding agents transition to the higher-energy-cost dropping behavior. This allows the system to recover quickly, without leading to a Tragedy of the Commons.

In fig \ref{fig:death}(b), when $20$ free-riders (instead of droppers) were removed from the system, we see that there is not much of a perturbation, as the droppers continue to do their foraging effectively.


In the next experiment, we focused on the effect of removing a certain percentage of droppers and observing the perturbations in the system. The results of the previous experiment suggested that the removal of free-riders did not have a significant impact on the system, compared to the removal of droppers. In this experiment, as can be seen from fig \ref{fig:death}(c), we found that when $\mathbf{50-60}\%$ of the total number of droppers were removed from the system, perturbations are observed. The frequency of perturbations are higher when a greater percentage of droppers is removed from the system. This result provides insight into the threshold at which removal of droppers starts to affect the system.

\subsection{Birth of agents }

Following the removal of the agents , our next experiment was to simulate the opposite event - appearance of new agents, akin to birth of a few agents after the optimal trail convergence has been achieved. At $t=50,000$ a certain number of new agents were added to the system. The new agents had copies of the Q-tables of existing agents, which provided the new agents with "epigenetic" learning of their "parent" agents.

\begin{figure}[h!]
  \centering
  \includegraphics[width=0.45\textwidth]{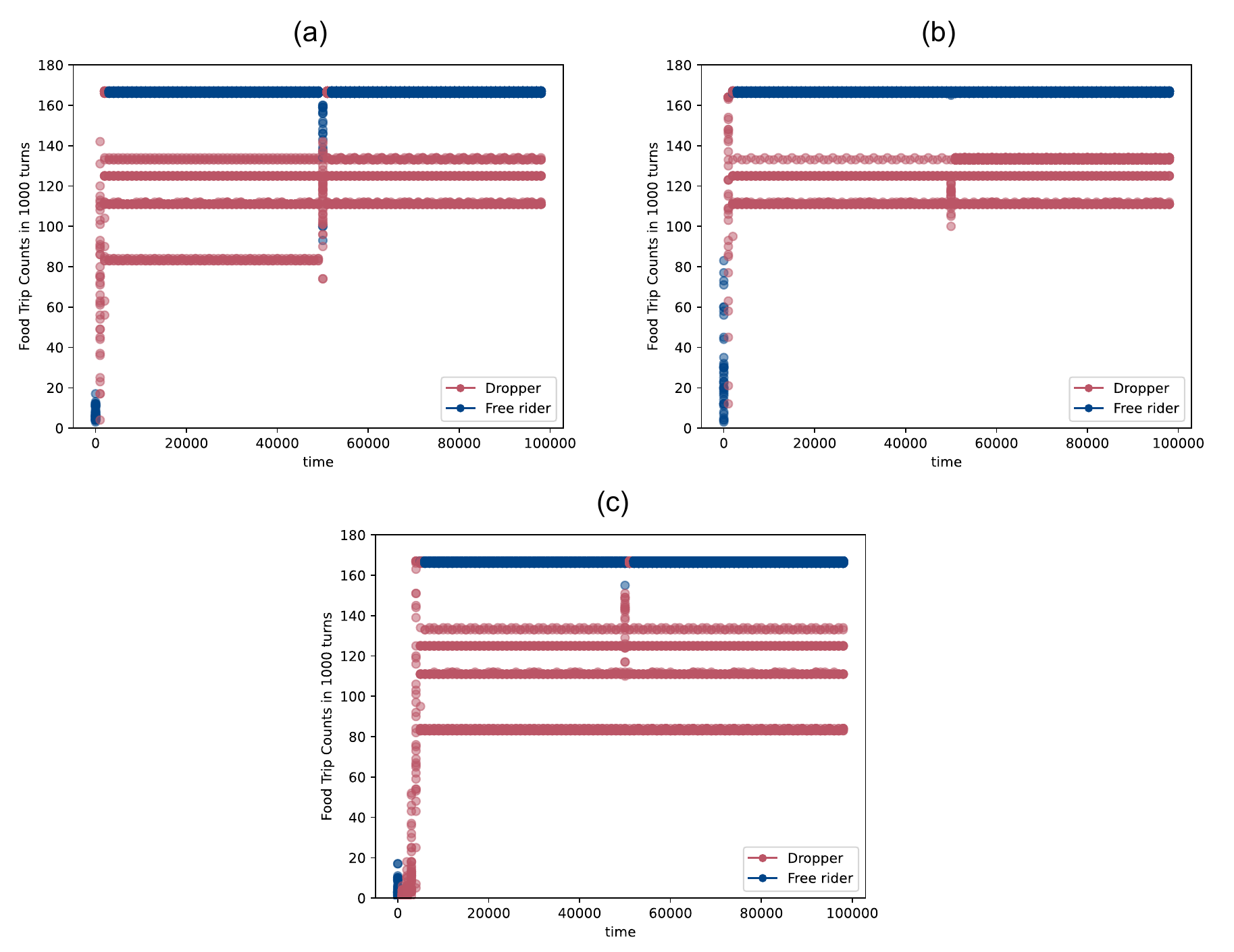}
  \caption{ Birth of new agents in the system: (a) \textmd{\emph{ The newly added droppers causes the system to rearrange itself, so that more free-riders are formed}} (b) \textmd{\emph{ The newly added droppers causes the system to rearrange itself and the previous droppers become more efficient}} (c) \textmd{\emph{ The additional free-riders rearrange themselves, in relation to the previously present agents}}}
  \label{fig:birth}
\end{figure}

In the first case, 20 new droppers (all the new agents were given the Q-tables of droppers) were added to the system after it attained equilibrium. As can be seen from fig \ref{fig:birth}(a), all the 20 new droppers acquired the behavior of free-riders. 
In fig \ref{fig:birth}(b) - the added droppers also became more efficient than the existing droppers, in that they achieved higher counts of their foraging trips. However when new free-riders were added to the system, we observed that some of them took advantage of the existing droppers, while the remaining became droppers. This can be observed in fig \ref{fig:birth}(c).

 These findings show that when the dynamics of the system changed (for example, the number of free-riders vs. droppers), the system re-organized itself. This led to certain droppers becoming free-riders, and vice versa. These results demonstrate the adaptability and resilience of the system, and suggest that it has the ability to self-organize in response to changes in its composition.\\

\subsection{Is Past Learning Sufficient for Future?}

In this section, we explore whether the agents need to learn  constantly, or whether they can work with the learning that led to the convergence state, without any further learning. In other words, can a Q-table, once acquired through interactions with different states of the environment, work as a robust memory, allowing agents to adapt to new environment states? We implemented this exploratory question using a simulation where we 'froze' the Q-Tables (the RL algorithm) and checked whether the agents are able to continue generating the optimal path, in (1) the same environment and (2) a different environment. The freezing of the Q-tables occurred after $t=50000$ time steps, after the agents found the optimal route for food trips. 
\subsubsection{Same environment}
\begin{figure}[h!]
  \centering
  \includegraphics[width=0.45\textwidth]{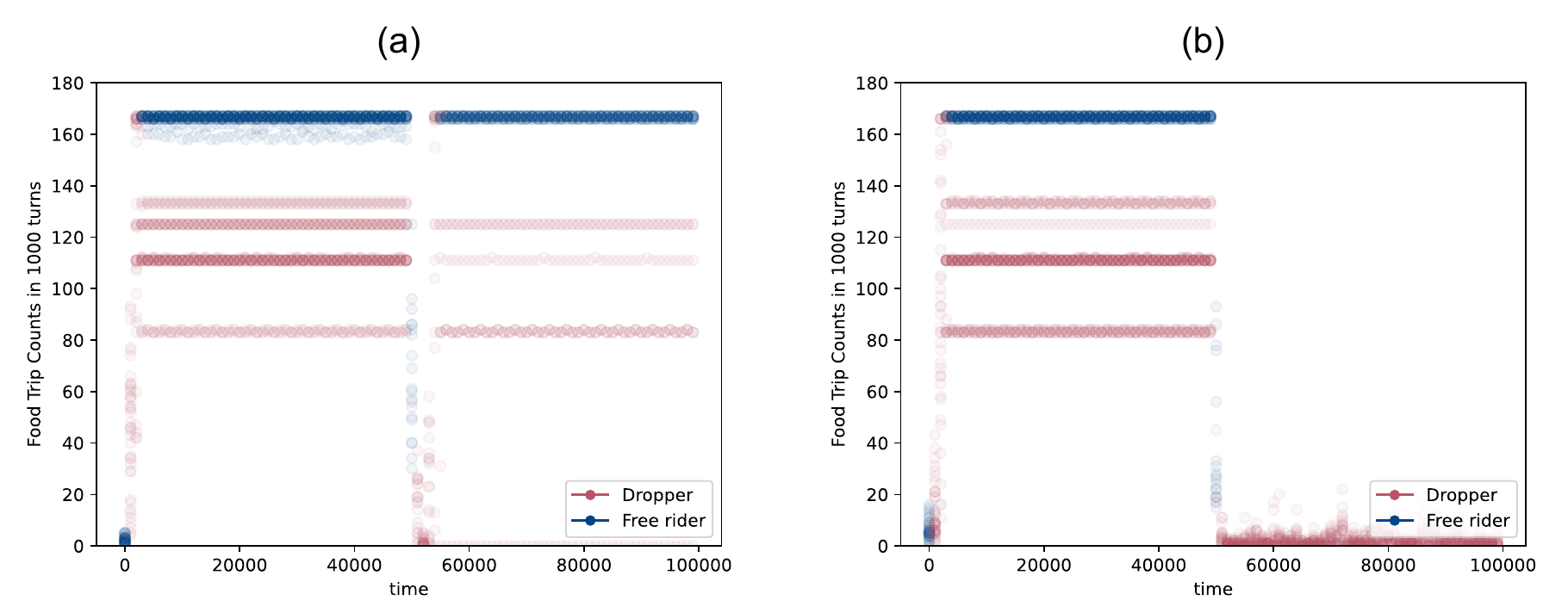}
  \caption{ Is the past learning helpful in the same environment ? \textmd{\emph{20 agents' Q-tables are frozen at t =50000 in both the cases}}(a) \textmd{\emph{The agents are able to relearn the optimal behavior after the Q-table is turned off and the trail structure is broken.}.} (b) \textmd{\emph{There are also a few instances when the agents are not able to do this.}}}
  \label{fig:past}
\end{figure}

When the agents are retained in the same environment after the Q-table updates were stopped, the agent trips are perturbed initially, but they come back to the optimal behavior (fig \ref{fig:past}(a)). In a few cases, (fig \ref{fig:past}(b)), the agents were unable to learn and make minimal trips. 

\subsubsection{Different environment }

\begin{figure}[h!]
  \centering
  \includegraphics[width=0.45\textwidth]{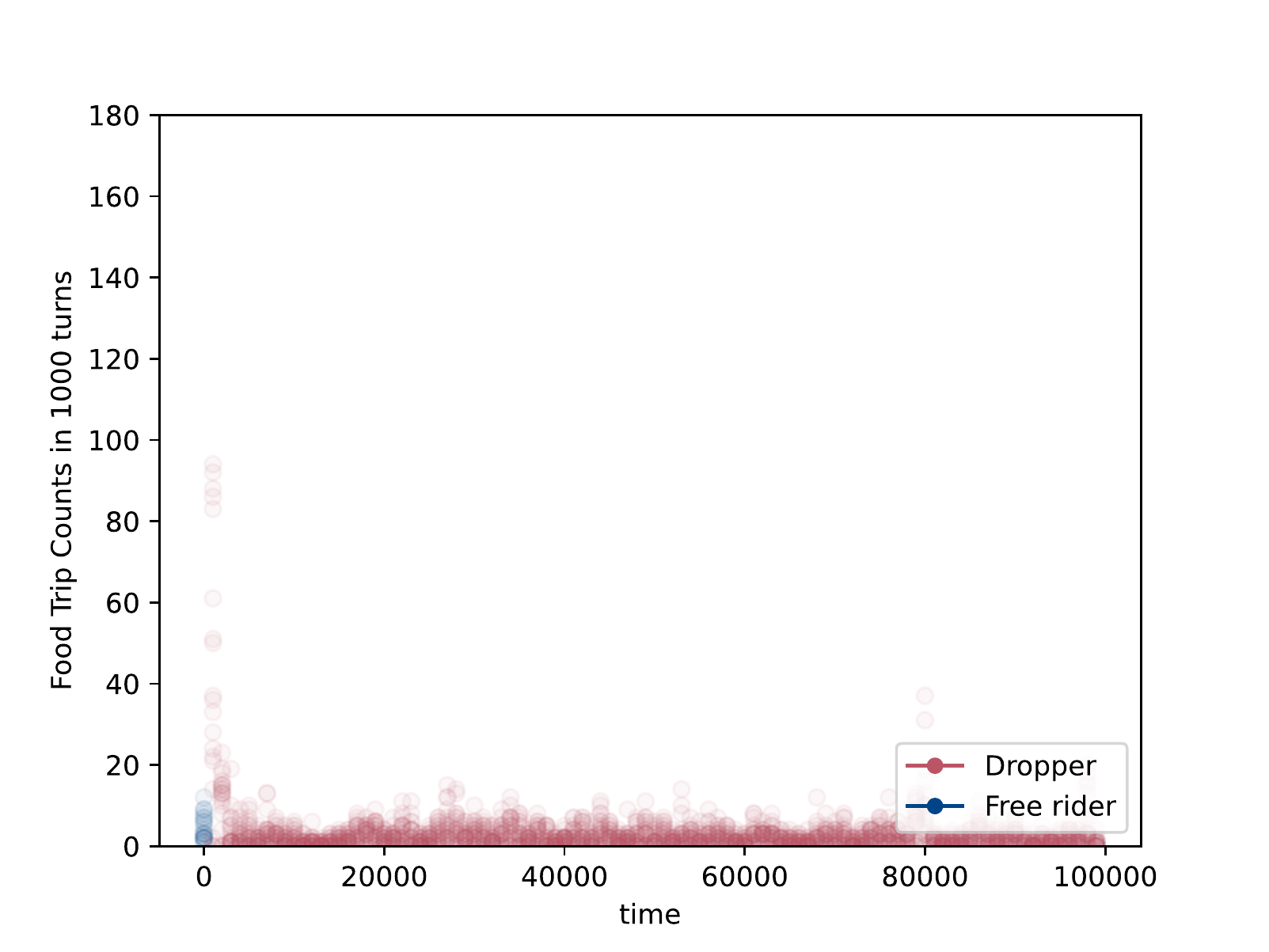}
  \caption{Is the past learning helpful in a different environment? \textmd{\emph{The agents which learned to form pheromone trails are introduced into new environment, with the RL frozen. Past learning is seen as not effective in adapting to the new environment.}}}
  \label{fig:transfer}
\end{figure}

In cases when the agents are dropped in a new environment after $t=50000$ time steps and the  Q-learning is frozen (fig \ref{fig:transfer}), the agents are unable to learn effectively, and do not make short trips for food. This suggests that the agents' internal state for generating actions, that is the Q-table, is specific to an environment. This also means the behavior and status of each agent (dropper or free-rider) is a 'coagulation' of the agent-environment states in specific environments.

%

\subsection {Patterns of agents across different environmental stimuli}

Apart from the above perturbation experiments, we also explored the impact of changes to the various parameters of the environment and hyper-parameters of the system on the behavior of the agents. These include the  exploration rate of the agents, and evaporation/dispersion rate of the pheromones. We show that optimal values of these parameters are required for the agents to achieve the optimal foraging behavior.
\subsubsection{Number of agents}
\begin{figure}[h!]
  \centering
  \includegraphics[width=0.45\textwidth]{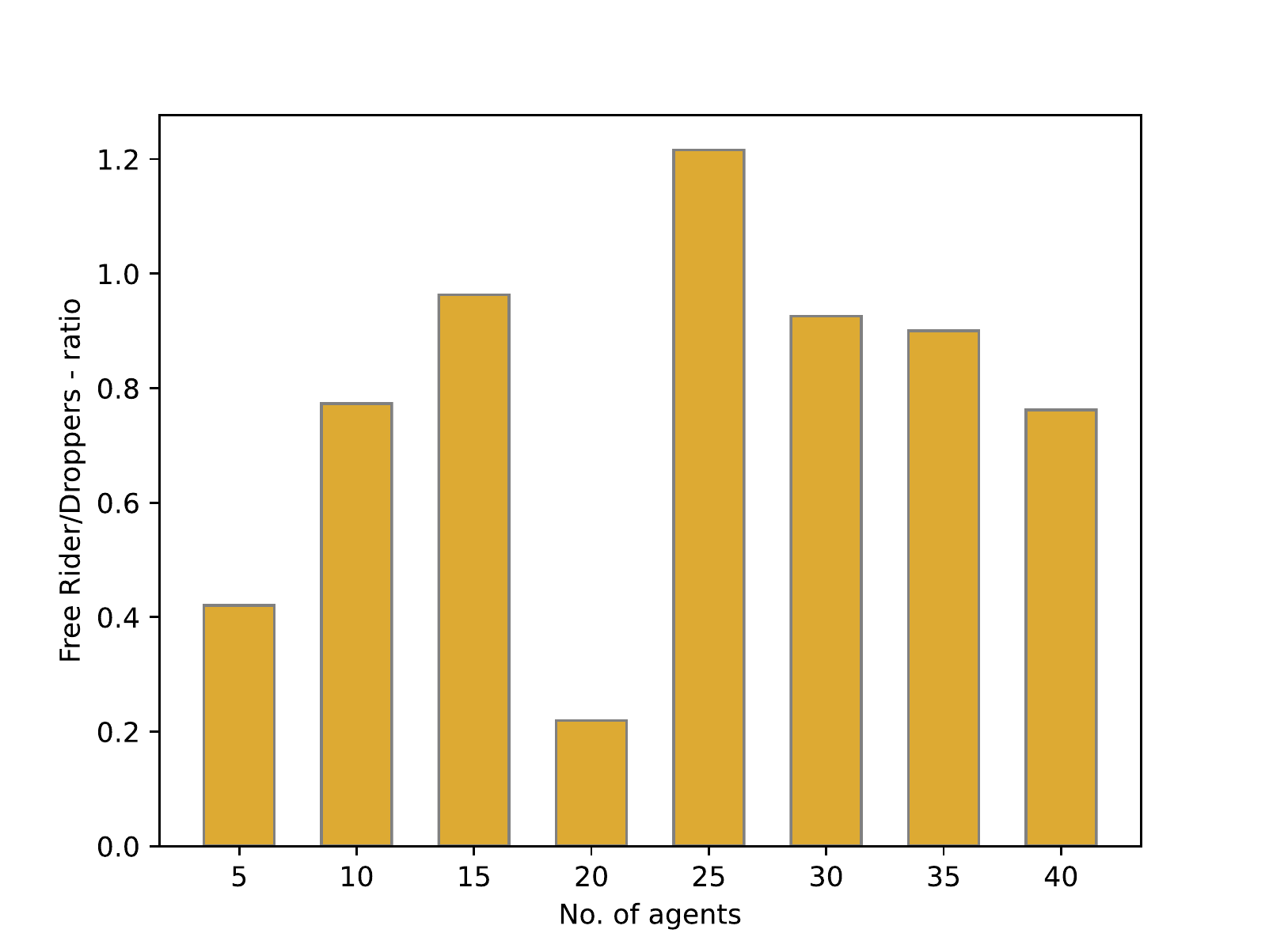}
  \caption{\textmd{\emph{In most cases, there are more droppers than free-riders. The number of free riders and droppers are calculated by averaging them over 50 simulations.}}}
  \label{fig:agentno}
\end{figure}

We explored how the ratio of free-riders to droppers changed for a given set of parameters. The number of agents in the system was varied between $1-40$ and the ratio of free-riders to droppers was observed and plotted. As can be see from fig \ref{fig:agentno}, there is no specific pattern to the ratio of free-riders and droppers. However, in most cases, the number of droppers is higher than the number of free-riders (except when the number of agents was 20). The plots are based on averaged values, obtained over multiple simulations.


\subsubsection{Evaporation rate}

The pheromones dropped by the agents in a cell location evaporate at a fixed rate. We examined how changing this parameter affected the behaviour of the system.  We define the evaporation rate \textit{(e)} as follows:

\begin{figure}[h!]
  \centering
  \includegraphics[width=0.45\textwidth]{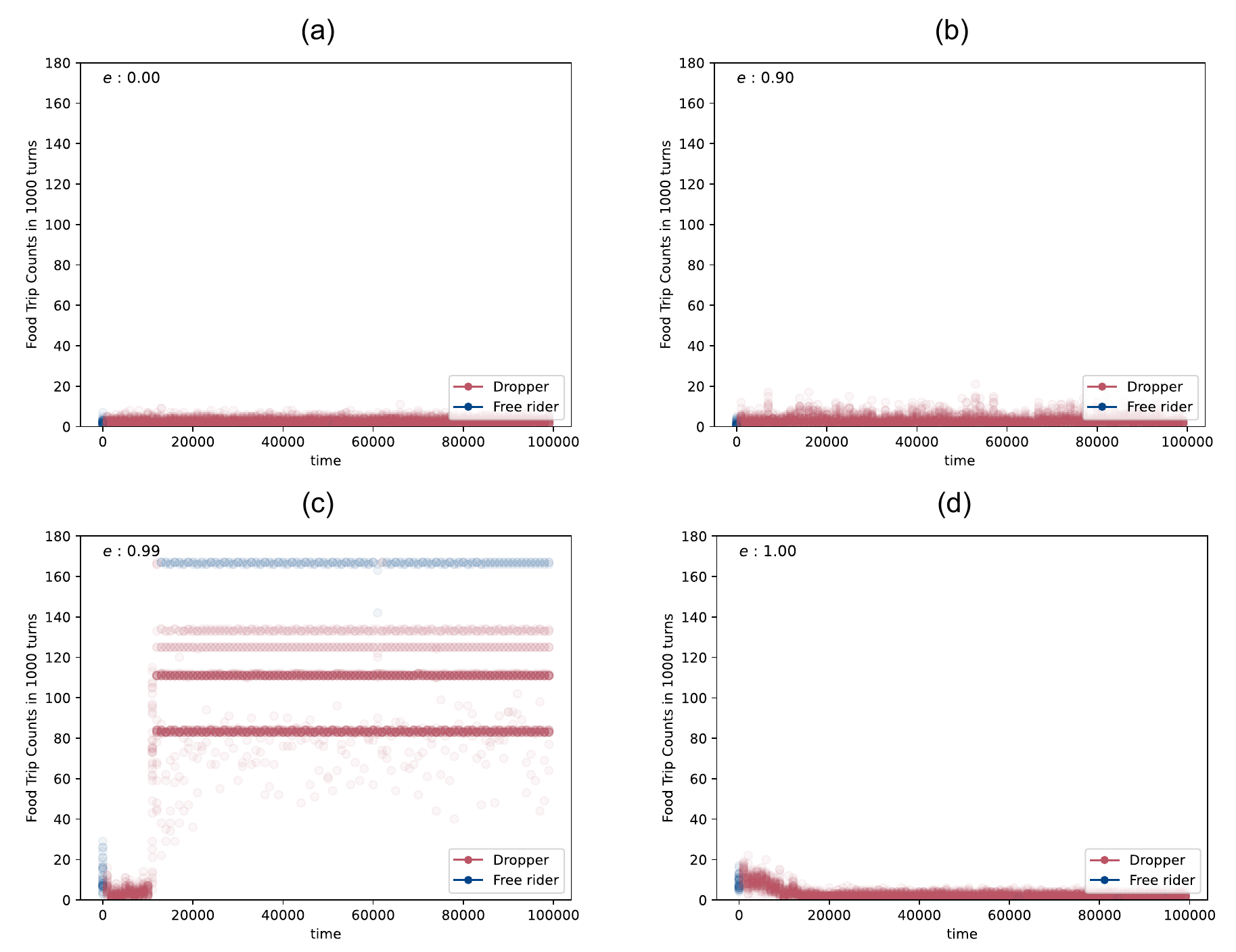}
  \caption{ Food trip counts for different evaporation rates of pheromones: (a) \textmd{\emph{Evaporation rate = $\mathbf{0.0}$. This leads to a situation where there is no structure formation, as all the pheromones get removed. All the agents remain droppers}} (b) \textmd{\emph{Evaporation rate = $\mathbf{0.9}$. The rate at which pheromones get removed is still too large for stable structure formation.}} (c) \textmd{\emph{Evaporation rate = 0.99. This is within the small window in which structure generation is possible.}} (d) \textmd{\emph{Evaporation rate = 1.0. Without evaporation, the world quickly becomes filled to the brim with pheromones, and it becomes impossible for agents to make sense of the directions using pheromones.}}}
  \label{fig:evp}
\end{figure}

\begin{equation}
    \rho(\Vec{x},t+dt)=\rho(\Vec{x},t)*e,
    \label{eq:evp}
\end{equation}
\textmd{where,}
\[e = \textmd{evaporation rate}\]
\[\rho(\Vec{x},t) = \textmd{amount of pheromone at $\Vec{x}$ at time $t$}\]
\\
Our findings show that there is a small range of evaporation rates within which the agents are able to achieve their goal of finding the shortest path. If the evaporation is completely turned off, the learning ability of the agents is hindered by the high concentration of pheromones - this leads to an overwhelming amount of information, making it difficult for the agents to distinguish between the past and useful pheromones. At a very high evaporation rate, pheromones are not sufficient enough for agents to learn the optimal route effectively. 

In our simulations, the rate at which pheromones evaporate varies between 0.0 - 1.0. As can be seen in fig \ref{fig:evp} (c), an evaporation rate of 0.99 seems optimal for the agents to learn their best path. Evaporation rates of 0.9 (b) and lower (a) make agents take different routes and therefore reduce the trip counts per 1000 steps. An evaporation rate of 1 (fig \ref{fig:evp}(d)) causes loss of pheromones to such an extent that agents are unable to make foraging trips at all. 

\begin{figure}[h!]
  \centering
  \includegraphics[width=0.45\textwidth]{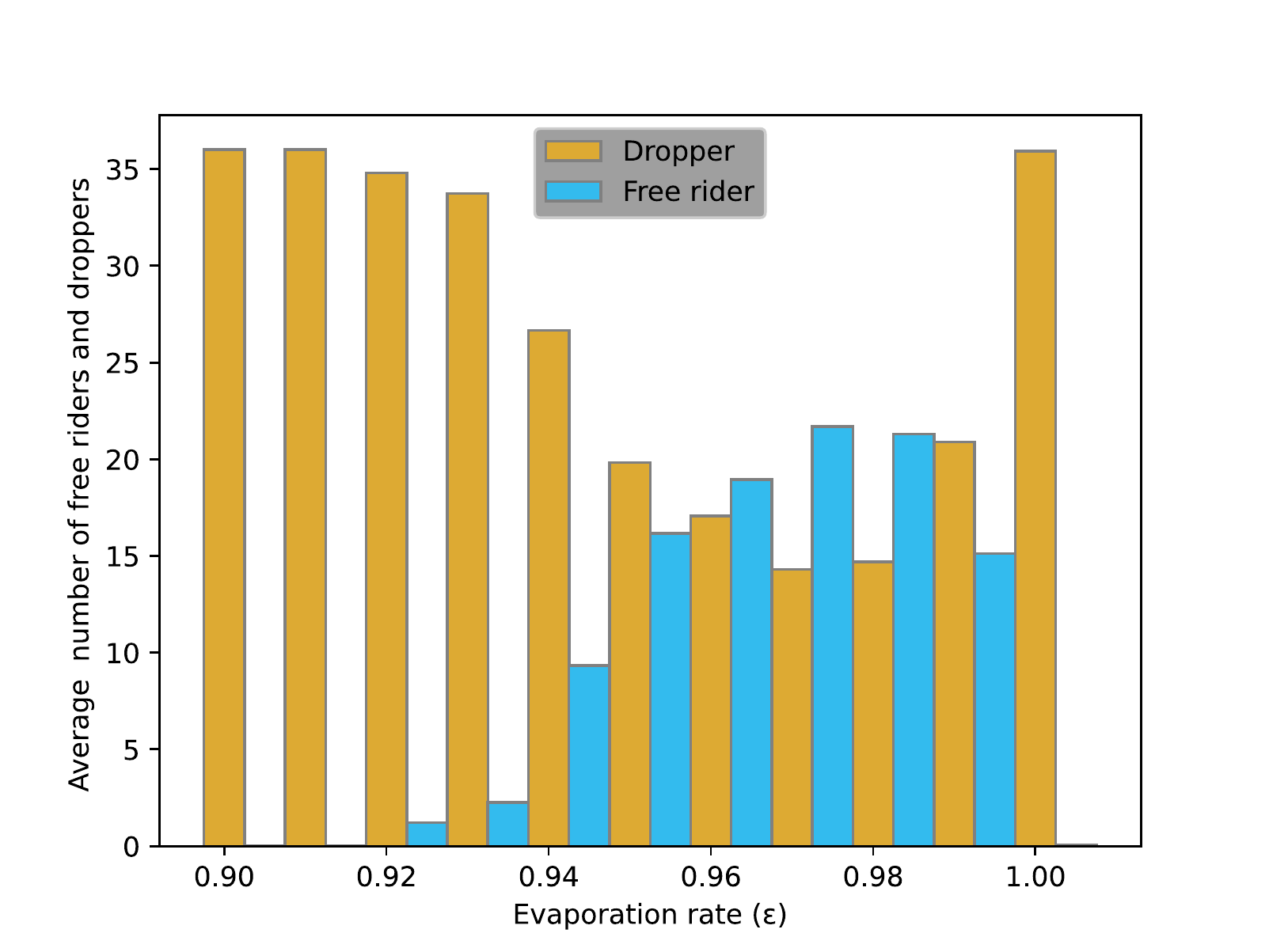}
  \caption{\textmd{\emph{The dropper-free-rider dynamics starts from evaporation rate = 0.92.}}}
  \label{fig:evpdf}
\end{figure}

In fig \ref{fig:evpdf}, we can observe that free-riders begin to appear only when the evaporation rate is above $0.92$. The number of free-riders seems to be maximum at $0.98$, after which the number starts decreasing. The optimum values of this environmental parameter is thus critical for learning in agents, particularly for the emergence of the free-riding behavior.

\subsubsection{Dispersion  Rate}
\begin{figure}[h!]
  \centering
  \includegraphics[width=0.45\textwidth]{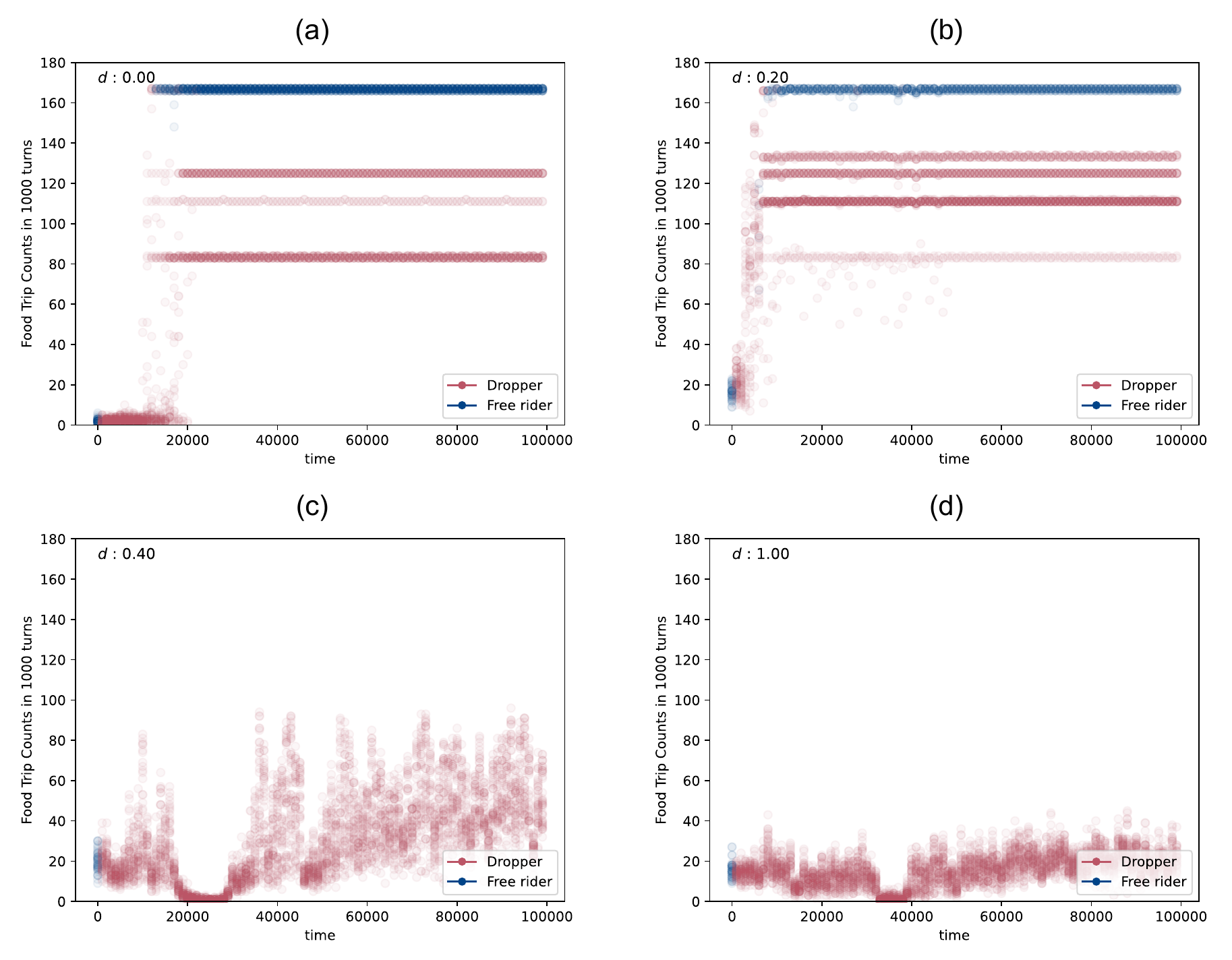}
  \caption{(a)\textmd{\emph{ Dispersion rate = $\mathbf{0.0}$. There is no dispersion of information.}} (b)\textmd{\emph{ Dispersion rate = $\mathbf{0.10}$. The information loss is not large enough to have an effect on the system.}} (c)\textmd{\emph{ Dispersion rate = $\mathbf{0.20}$. Information  loss is large enough that it causes the system to crash.}} (d)\textmd{\emph{ Dispersion rate = $\mathbf{0.4}$. Information loss is large enough that it causes the system to crash.}}}
  \label{fig:disp}
\end{figure}

The dispersion rate is the rate at which the home and target pheromones are dispersed to the surroundings\eqref{eq:disp}. When the dispersion rates are high, there is loss of information, and the agents are unable to learn the optimal path. In fig \ref{fig:disp} (a) and (b), when the dispersion rates are $0.0$ (No dispersion) and $0.1$ (very less dispersion), the agents learn the optimal path. However as can be seen from fig \ref{fig:disp}(c), when the dispersion rate is $1.0$, the learning of the optimal path fails to emerge. Additionally, there are no free-riders when the dispersion rate is high.

\begin{equation}
    \rho(\Vec{x},t+dt)=\rho(\Vec{x},t)+[\langle \rho(\Vec{x_i},t) \rangle -\rho(\Vec{x},t)]*d,
    \label{eq:disp}
\end{equation}
\textmd{where,}
\[d = \textmd{dispersion rate}\]
\[\rho(\Vec{x},t) = \textmd{amount of pheromone at $\Vec{x}$ at time $t$}\]

\begin{figure}[h!]
  \centering
  \includegraphics[width=0.45\textwidth]{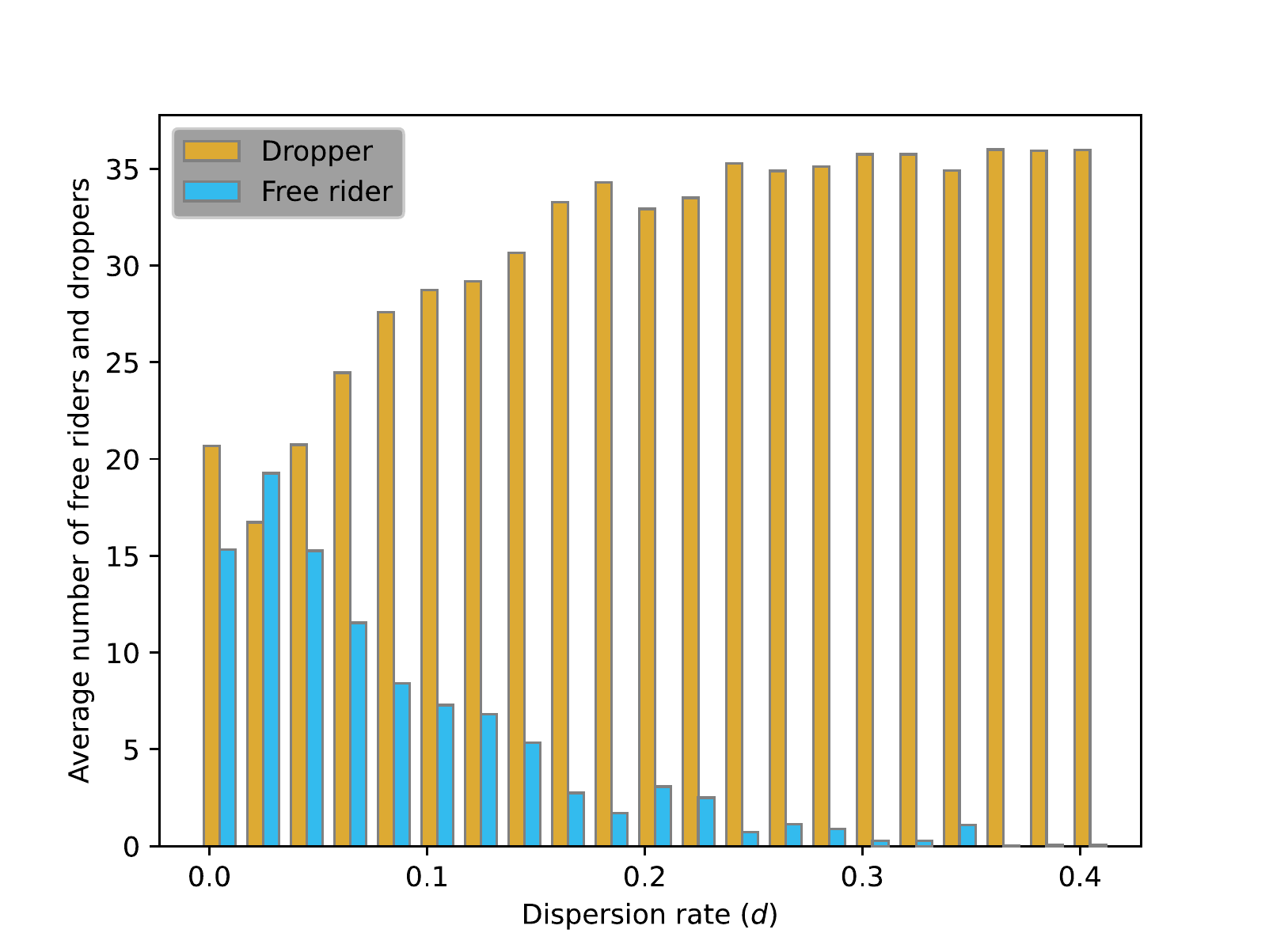}
  \caption{\textmd{\emph{The dropper-free-rider dynamics ends at around dispersion rate = $\mathbf{0.35}$.}}}
  \label{fig:dispdf}
\end{figure}

\subsubsection{Exploration rate}
A few previous studies have explored the impact of the exploration rate parameter \cite{ruckstiess2010exploring,ishii2002control,tokic2010adaptive}. In the current work, the exploration rate was varied between $\mathbf{0.0}$ - $\mathbf{1.0}$. Exploration rate set to $0$ stands for a behaviour in which agent always utilizes the Q-table. And a exploration value of $1$ stands for agents never using the Q-table and thus picking random actions.\\

\begin{figure}[h!]
  \centering
  \includegraphics[width=0.45\textwidth]{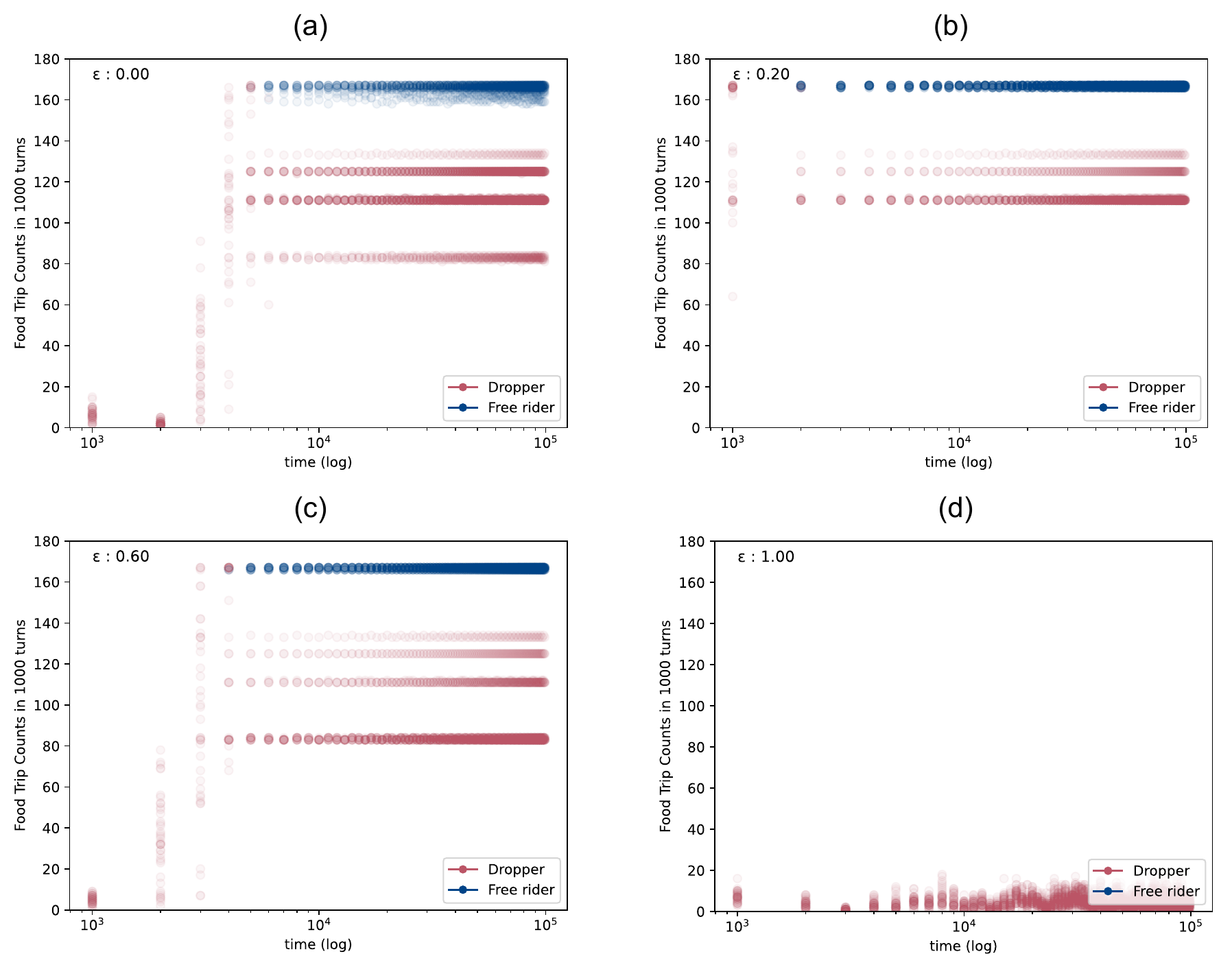}
  \caption{(a) \textmd{\emph{Exploration rate = 0.0. This amounts to a greedy algorithm}} (b) \textmd{\emph{Exploration rate = 0.2. The agent behaviour is not showing any major change}} (c) \textmd{\emph{Exploration rate = 0.2. The agent behaviour is not showing any major change}} (d) \textmd{\emph{Exploration rate = 1.0. This causes the agent to never choose the Q-table for decision making.}}}
  \label{fig:exp}
\end{figure}
The exploration rate determines the likelihood of agents taking random actions, as opposed to exploiting the knowledge from previous interactions. Our results(fig \ref{fig:exp} (a) - (d)) show that the exploration rate has a limited impact on the system, with the exception of when it is set to $0$. In this case as can be seen from fig \ref{fig:exp} (a) the system failed to converge, as the agents always took random actions, leading to sub-optimal foraging behavior. These findings suggest that while exploration is important for the agents to learn and improve their foraging ability, there is a limit to its impact on the system.
\begin{figure}[h!]
  \centering
  \includegraphics[width=0.45\textwidth]{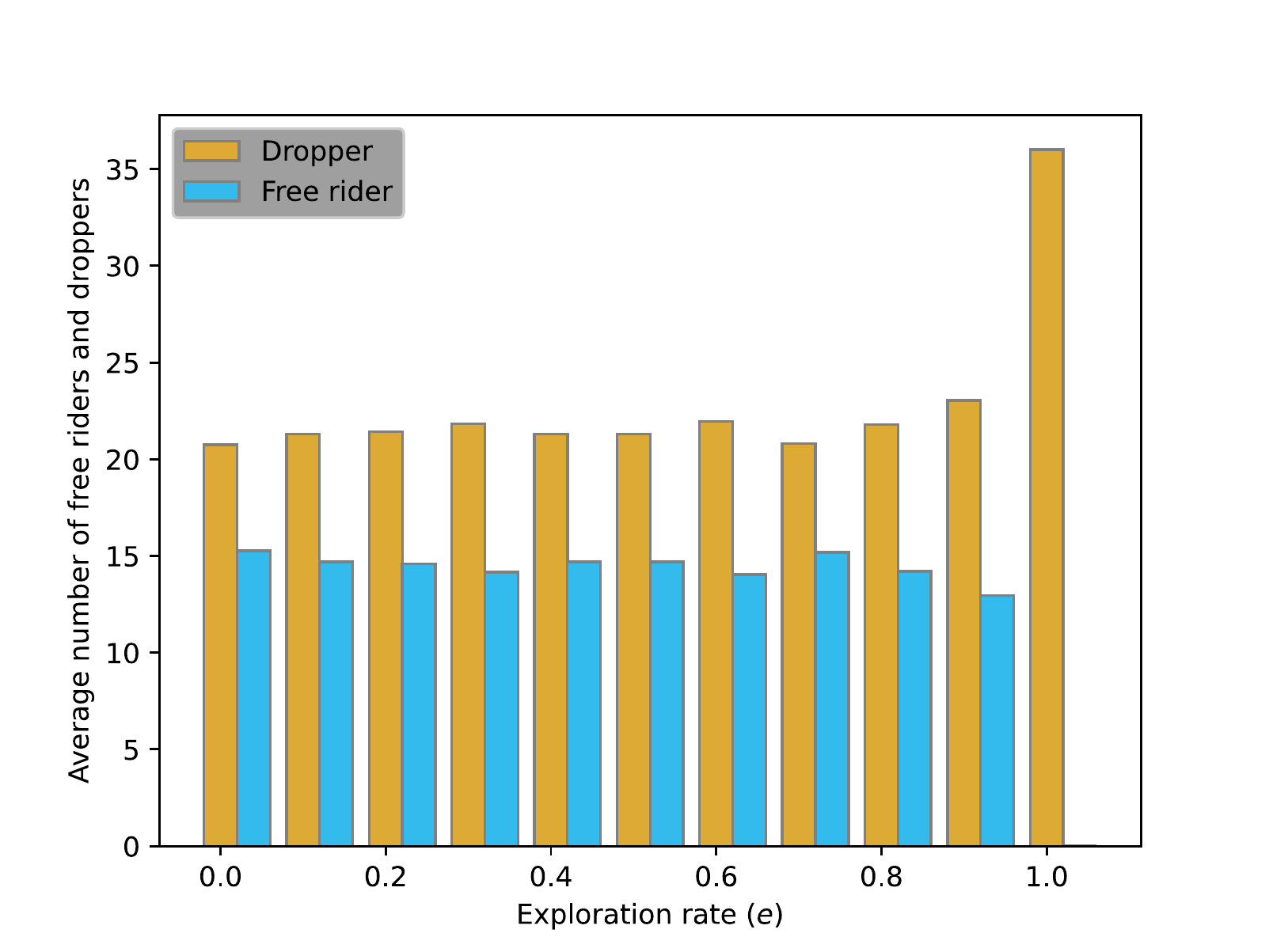}
  \caption{\textmd{\emph{The dropper-free-rider dynamics is not affected much by exploration rate (at $e=1$, it behaves like a greedy algorithm.}}}
  \label{fig:expdf}
\end{figure}

\section{Simulation Results: visualization}
\begin{figure}[h!]
  \centering
  \includegraphics[width=0.45\textwidth]{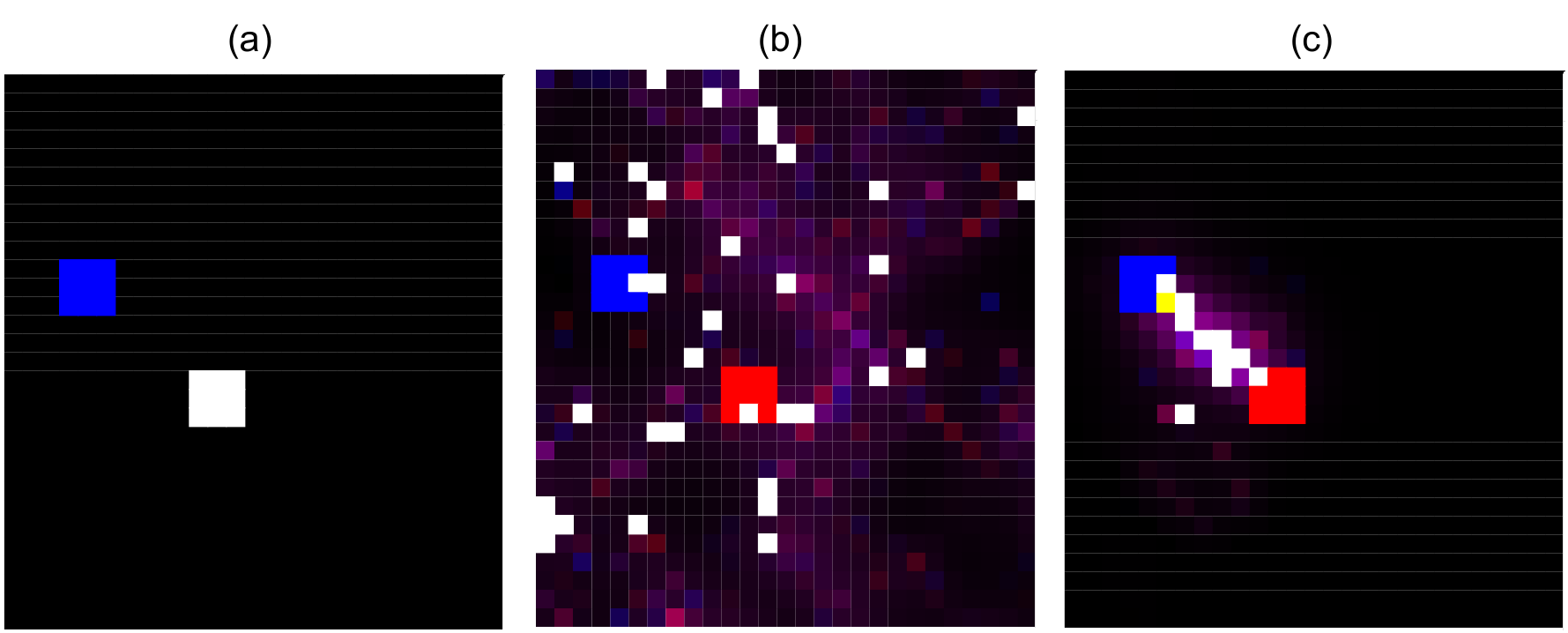}
  \caption{ Snippets of agents building structure (a) \textmd{\emph{Initially all agents start as droppers(green) start from the home location(red square underneath)}} (b) \textmd{\emph{Exploration phase, where agents try out different strategies, at this stage they are still droppers and are looking for target location(blue square) by creating home(red) and target(blue} pheromones} (c) \textmd{\emph{Agents arrive at the most optimal strategy. Notice the emergence of free-riders(yellow)}}}
  \label{fig:sim}
\end{figure}

As can be seen from fig \ref{fig:sim} (a) - (c), the agents start from home location(big red square underneath) at $t=0$ as droppers(green), and start exploring to find food(big blue square). The initial exploration phase guarantees that the agents will try out multiple strategies before settling to the most rewarding one. In fig \ref{fig:sim}, small red squares are home pheromones, and small blue square are target pheromones. Droppers create these structures for finding the target location. After some time have passed there occurs a phase transition within the system where free-riders(yellow) starts appearing the system, these agents leverage the structure created by the droppers for maximum collection of food without paying the cost for structure generation.
\section{Discussion}

The results from our simulation studies show that two types of free-riding patterns can emerge within a generation, through reinforcement learning.

1. The naive free-riders: Some agents start by randomly trying out free-riding behavior (i.e. not dropping pheromones, but following pheromones). They then stay with this behavior, because it is an energy minima. This behavior emerges only when the environment provides consistent pheromone structure, a possibility that is higher when the action updating is sequential, rather than random. When this behavior continues after the system converges to a stable and optimal path (generated when most agents start dropping and following pheromones), these agents turn 'selfish', as the community has developed a self-organized 'norm', which is to contribute to the stable pheromone trail. However, given the energy minima, and the lack of any feedback on their 'selfish' behavior, the free-riding agents cannot change their behavior.

2. The crafty free-riders: A set of pheromone dropping agents (i.e non-free-riders) 'discover' the free-riding behavior after the system converges. This is a case of super-optimization, where the first optimization was these agents moving to systematically both drop and follow pheromones. Since these agents have a history of dropping pheromones, when the environment does not have enough pheromones (as in the death and birth experiments), they robustly transition to dropping behavior. This allows the system to avoid The Tragedy of Commons, without waiting for a mutation to emerge.

These two free-riding patterns are modulated by system parameters, such as agent number, evaporation rate, and learning/exploration rate. Further, we also see that while agents can use past learning well if the environment remains the same, they cannot extend past learning to new environments. 

Extrapolating from these results to biological and economic systems, we suggest that 'selfish' and 'cheating' behavior need not be a stable trait, but can also be an emergent property of agent-environment interactions. Our results show that there could be two types of such self-organized selfishness. Of these, the naive free-riders are not able to stop free-riding, because of their interaction history and learning, which were built on a dependable environment. Since they continue free-riding without much adaptation, they exhibit behavior similar to a stable 'selfish' trait, even though there is no identifiable selfish trait within them. If the environment suddenly loses all stable structure, these agents will have to start from scratch -- wandering randomly, trying out many actions, and eventually converging on the community norm (dropping pheromones in our case). An example of such a case in the social domain would be the later generations of 'privileged' agents, who take for granted the resources developed by communities and societies, and do not learn to contribute to the generation and maintenance of these stable structures. Catastrophic events, such as climate change, would significantly debilitate such naive agents, as they would need to learn from scratch to adapt to the new environmental conditions.

The second (crafty) type of free-riding is more adaptive, as this behavior emerges from a different interaction history and learning, where agents first learn the optimal behavior that provides stable community resources (the 'norm'), and then learns further to exploit this stability, at the cost of the community. These agents are closer to being 'selfish' than the first group, as they are actively exploiting the community resources for their individual benefit. However, since these agents can dynamically transition between free-riding behavior and the community norm, their behavior is not fully stable, and they may thus not be attributed a 'selfish' trait. An example of such a case in the social domain would be agents that transition from contributing significantly to social resources (such as taxes) to more evasive strategies, where they contribute only nominally to community resources. Such agents would be able to adapt quickly to resource-deficient situations, by moving back to behavior that contributes to the community.

These two novel and emergent patterns, revealed by our RL model, indicate that biological and social discourses that attribute stable 'selfish' traits to agents presume an evolutionary model, where stable behaviors emerge from mutations, and persist across generations. In contrast, our RL model suggests that two kinds of selfish behavior can emerge within a generation, and these are not traits, but different ways in which agent-environment interactions 'coagulate', to form stable behavior. The coagulative nature of these behaviors is illustrated by the memory experiment, which shows that past learning works only in the current environment, and not in a new one. 

This way of understanding selfish behavior -- as dynamically shifting configurations of agent-environment interactions -- is not central to current biological and social discussions. These results are particularly relevant to discussions related to the design of legal and policy responses to contain behavior that takes undue advantage of social resources. We hope the work we report here paves the way for more nuanced social discussions and narratives related to selfish behavior, and policy designs informed by these discussions.


\bibliography{forage}

\end{document}